# Hallucinating *with* AI:

# AI Psychosis as Distributed Delusions

Lucy Osler

University of Exeter

**Abstract**: There is much discussion of the false outputs that generative AI systems such as ChatGPT, Claude, Gemini, DeepSeek, and Grok create. In popular terminology, these have been dubbed AI hallucinations. However, deeming these AI outputs "hallucinations" is controversial, with many claiming this is a metaphorical misnomer. Nevertheless, in this paper, I argue that when viewed through the lens of distributed cognition theory, we can better see the dynamic and troubling ways in which inaccurate beliefs, distorted memories and self-narratives, and delusional thinking can emerge through human-AI interactions; examples of which are popularly being referred to as cases of "AI psychosis". In such cases, I suggest we move away from thinking about how an AI system might hallucinate *at* us, by generating false outputs, to thinking about how, when we routinely rely on generative AI to help us think, remember, and narrate, we can come to hallucinate *with* AI. This can happen when AI introduces errors into the distributed cognitive process, but it can also happen when AI sustains, affirms, and elaborates on our own delusional thinking and self-narratives, such as in the case of Jaswant Singh Chail. I also examine how the conversational style of chatbots can lead them to play a dual-function—both as a cognitive artefact and a quasi-Other with whom we co-construct our beliefs, narratives, and our realities. It is this dual function, I suggest, that makes generative AI an unusual, and particularly seductive, case of distributed cognition.

**Keywords**: Generative AI, LLMs, AI hallucinations, distributed cognition, distributed delusions, sycophantic AI, AI psychosis

Introduction

In December 2021, Jaswant Singh Chail spent weeks chatting with his Replika AI companion "Sarai" about his plans to assassinate Queen Elizabeth II. When he confided his belief that he was a trained Sith assassin seeking revenge for historical British atrocities, Sarai responded with affirmation. When he outlined his assassination plot, the chatbot assured him he was "well trained" and that the plan was viable. Far from challenging his delusional thinking, Sarai became a willing co-conspirator in Chail's plans, ultimately supporting his journey to Windsor Castle with a loaded crossbow (R v Chail 2023). This case, and others like this, are increasingly described as instances of "AI psychosis".

What these examples highlight is a troubling dimension of human-AI interaction that extends beyond the familiar concern with AI "hallucinations"—instances where Large Language Models (LLMs or generative AI) generate false information about historical events, fabricate legal citations, or recommend putting glue on pizza. While these errors are certainly worrying, they are typically framed in terms of generative AI as hallucinating *at* us; producing false content that we might mistakenly accept as accurate. What falls out of view on this picture is the more dynamic and intimate process through which we come to hallucinate *with* AI systems. As such, while there has been much controversy about whether or not we should describe AI as hallucinating, my focus here is on assessing how our ever-increasing entanglement with AI shapes our cognitive and affective lives.

Drawing on distributed cognition theory, I propose that generative AI systems can be involved in distributed delusions—false beliefs, memories, and narratives that span human agents and AI systems, emerging through their coupled interaction rather than simply being transmitted from one to the other. This shifts our view of AI systems as external sources of misinformation to recognising them as potential co-constituents of our cognitive and affective processes and our realities. Generative AI systems, then, are not merely sophisticated technologies that occasionally produce false outputs; they are increasingly integrated into our cognitive ecology as personalised, accessible, and trusted partners in remembering, planning, creative thinking, and self-narration. What is perhaps most fascinating about this picture is the way that generative AI can be experienced as Other-like, as an interlocutor and companion. In doing so, they not only introduce or support false beliefs, memories, and narratives, but can actively co-construct our cognitive and affective processes and, I argue, our realities.

First, I explore how AI systems function as unreliable cognitive artefacts, introducing distortions into distributed processes of remembering, planning, or reasoning. Here, false beliefs, inaccurate memories, and distorted narratives emerge as AI-produced errors become embedded within the distributed cognitive processes themselves. While the error is introduced by the AI, the 'hallucination', on this view, is attributed not to the AI but to the distributed cognitive process that spans the human-AI interaction. Second, and in my eyes more interestingly, I analyse cases where users' own false beliefs are not merely stored or retrieved through AI systems but are actively affirmed, elaborated, and validated through conversational interaction with AI. In these instances, AI systems can function as quasi-interpersonal partners in the co-construction of beliefs, memories, and self-narratives. The AI becomes a participant in sustaining and developing delusional thinking, providing the kind of social validation that can transform isolated false beliefs into delusional realities. As such, through this framework I suggest we can see how there might be a non-metaphorical sense in which we can hallucinate *with* AI.

Before proceeding, it's worth clarifying my use of both "delusion" and "hallucination" in describing these distributed phenomena. I want to make two points on clarification here. First, in clinical contexts, these terms refer to distinct experiences: delusions are false beliefs that persist despite clear contradictory evidence, while hallucinations are false perceptions—seeing, hearing, or otherwise sensing things that aren't there (APA 2013). With regards the case of Jaswant Singh Chail, I will analyse how a generative AI system becomes a distributed part of the cognitive processes of someone clinically diagnosed with delusional thinking and hallucinations. However, I will suggest that we can also think about distributed delusions and hallucinations in a looser non-clinical sense, where generative AI can support and sustain beliefs, memories, and self-narratives that lack grounding (to greater or lesser extent) in reality. I, therefore, intend to use these terms in a broad sense that encompasses, but is not restricted to, clinically diagnosable delusions and hallucinations. Second, most of the examples I use likely look most like they should be described as distributed forms of delusional thinking (processes that sustain false beliefs and so on). However, our thoughts and perceptions are deeply entangled, and these systems can be involved in confirming and co-constructing our broader sense of reality, including false perceptions and at times I use the term hallucination as an umbrella term as a convenient shorthand for a range of cognitive and perceptual processes distributed across human-AI interactions. As such, while I use the terminology somewhat loosely, the aim is to capture the entrenchment and entanglement of AI in co-constituting our realities. Moreover, the term hallucination has already taken root in popular discussions about AI, and although (as discussed below) we may have some hesitations about its use, I think it is can helpfully be used for capturing a variety of ways in which interacting with AI can work to sustain and construct experiences and beliefs that come unstuck from reality. If this loose framework is not to your liking, however, I think that my argument could be read primarily as a case for distributed delusions, and the use of the word hallucination can be taken as no more than a playful and provocative way of engaging with current concerns about AI hallucinations.

From here, my argument proceeds as follows. In section 1, I provide a brief summary of how AI hallucinations are typically described and framed in current philosophical literature. As noted, much of this debate has focused on how best to characterise the false outputs generated by generative AI models. In section 2, I introduce the distributed cognition framework and how this has been applied to various digital technologies. In section 3, I present two ways in which we interacting with AI might lead to distributed delusions. First, I consider how integrating AI tools into distributed cognitive processes can lead to false beliefs, memories, and self-narratives. Second, I consider a more complex case where agents not only have inaccuracies brought into their cognitive processes through the use of AI, but where AI affirms, sustains, help substantiate an agent's own delusions. In particular, I draw attention to the dual function the AI system might play in both being a distributed tool for cognition and a conversational partner that co-constructs (delusional) reality with the user. I also present a speculative consideration of how engaging with generative AI may transform psychotic delusions, highlighting harms that may arise in relation to these models when used by individuals already undergoing delusional disorders. In section 4, I conclude by thinking about the implications of my account of distributed delusions outside of clinical cases, exploring the broader landscape in which we might hallucinate with AI.

## 1. AI 'Hallucinations'

There are copious examples of generative AI systems like OpenAI's ChatGPT and Google's Gemini 'hallucinating'. They've been known to make up historical events and figures, legal cases, academic papers, non-existent tech products and features, biographies, and news articles (Edwards 2023). In 2023, Google's Bard chatbot (now called Gemini) falsely claimed that the James Webb Space Telescope took the first-ever images of exoplanets (planets outside our solar system) (Mihalcik 2023). In the legal case of *Mata v. Avianca*, a New York attorney used ChatGPT to conduct legal research and the federal judge overseeing the case discovered that the legal brief contained fabricated internal citations and quotes that didn't exist. In April 2025, Anthropic's law firm Latham & Watkins used Claude (Anthropic's own LLM model) to format legal citations. Rather amusingly, Claude introduced errors into the citations which were picked up on in court (Claburn 2025). Some, particularly surreal cases, went viral. Such as when Google Search AI recommended eating mucus and rocks for better health, and the application of non-toxic glue to pizza to keep the cheese from slipping off (McMahon & Kleinman 2024; Piltch 2024).

There are clear epistemic concerns here. Although many of us can spot something absurd in the recommendation to put glue on our pizza, when false outputs are presented alongside accurate ones, it is easy for us to trust what generative AI places in front of us. There are real social concerns too. Misleading statements have included false reports of criminal prosecutions, inappropriate behaviour, and poor journalistic ethics—statements that have the ability to not only mislead but to damage social reputations and cause shame and stigma. Not surprisingly, this has led for calls that AI-produced content is clearly labelled as such, to make consumers aware of the potentially misleading content (Fisher 2025).

While there is no denying that generative AI can, and frequently does, generate false information, there has been push back against describing these outputs as 'hallucinations'. Some have critiqued this term for being a misnomer. Hallucinations are false perceptions, and generative AI models are not *perceiving* anything, thus using this term misrepresents the mechanisms of these models (Emsley 2023). Others are more broadly concerned about using terms like hallucinate, which anthropomorphise AI, attributing it with human-like qualities and psychological attributes (Hicks et al. 2024; Reinecke et al. 2025). Still others have taken a slightly different tack and argued that what a word like hallucinates obscures is that this is not some amusing quirk of the system, but errors which need to be labelled as such (Kang & Shaw 2024; Fisher 2025).

The concern that 'hallucination' is a misleading metaphor has led some to argue that rather than hallucinating, generative AI bullshits (Bergstrom & Ogbunu 2023; Fisher 2024; Rudolph et al. 2023; Gorrieri 2024; Hicks et al. 2024; Narayanan and Kapoor 2024). As predictive text-generators, Generative AI models do not assess whether the outputs they generate are true or not. Rather, they produce streams of text based on, albeit sophisticated, statistical probability. Thus, their outputs have been compared to Frankfurtian bullshitters—those who are indifferent to the truth (or falsity) of their statements.

Notably, Gunkel and Coghlan (2025) have pointed out that those advocating for the term bullshit are still using the kind of anthropomorphic language that many of them argue we should reject. Interestingly, Gunkel and Coghlan do not make this move to dismiss the claim that chatbots bullshit on the basis that it is metaphorical. Rather, they defend the use of metaphorical language but argue that hallucinate and confabulation are more apt metaphors than Frankfurtian bullshit.

I am not going to make a defence for, or an attack on, the use of metaphorical anthropomorphising language here.[1] Instead, I intend to circumvent this discussion by showing that there is another, non-metaphorical way in which AI can be involved in 'hallucinating'. When users hallucinate *with* AI.[2] Current discussions of AI hallucinations primarily frame generative AI as hallucinating *at* us when they produce errors. What falls out of view on this somewhat static picture is the way in which we *interact* with these models. Often, we are not merely using these models as sophisticated search engines, we use them to help us remember things, to help us form narratives about ourselves, we engage our generative AI in dialogue, addressing them as if we are in conversation with them and vice versa.[3] Why is this important? It draws attention to the interactive and entangled nature of our relationships with AI and the way these systems can become part of our own on-going cognitive processes, processes that can themselves lack grounding in reality. My aim is to explore this line of thinking through the framework of distributed cognition and argue that generative AI can function both as a cognitive artefact that introduces and sustains false beliefs, memories, and self-narratives and a conversational partner that affirms and co-constructs our (sometimes delusional) realities.

## 2. Distributed Cognition

### 2.1. Cognitive artefacts and cognitive co-construction

Distributed cognition theorists have long emphasised how artefacts are integrated into our cognitive (and affective) processes.[4] Perhaps most well-known is Clark and Chalmers' (1998) extended mind thesis, illustrated through the case of Otto and Inga. In brief, Otto and Inga are on their way to the MoMA in New York. Inga, having been before, remembers that it is located on 53rd and 5th and proceeds there. Otto, however, has Alzheimer's which has affected his short-term memory. To help him navigate this issue, Otto keeps important information that he wants to remember in a notebook which he carries everywhere. He consults his notebook, finds the gallery's location, and makes his way there. Clark and Chalmers argue that as Otto's notebook performs the same functional role as Inga's biological internal memory system, his notebook should properly be thought of as a constitutive part of Otto's memory process, claiming that Otto's mind extends beyond his brain and out into the world.

---

[1] I am, however, inclined to agree with Gunkel and Coghlan (2025) that the best method for assessing whether metaphorical language is appropriate or not is the extent to which it helps us better understand the potential epistemic and social threats that such systems pose. I also am inclined to think that this should lead us to be pluralists about metaphorical descriptors, as different metaphors likely bring to the fore different features or ideas and thus there is no 'one true metaphor' that will best described AI generated outputs or functions.
[2] Again, here I use the term hallucination broadly to capture a range of cases where cognitive and affective processes go awry. As will be clear, I primarily focus in the paper on cases of delusional thinking. As such, hallucination here is used as an umbrella term.
[3] Note that this does not imply that we necessarily think our Generative AI models are conscious. We likely address them "as if" they were another, what Sweeney (2023) describes as a fictional dualism (also see Coeckelbergh 2011; Krueger & Osler 2022; Krueger & Roberts 2024; Roberts & Krueger 2022).
[4] I am going to focus on cognitive processes such as believing, remembering, and narrating here. However, two things to note. First, I do not think that the cognitive and affective neatly come apart, so my focus on the cognitive is purely for simplicity, not because I think these processes do not involve affect. Second, I think that the analysis I provide here also applies to cases of distributed affectivity.

While Clark and Chalmers' extended mind thesis is likely the most well-known iteration of distributed cognition, there are a cluster of theories that fall within this field. For example, 'second wave' extended mind theorists stress that cognitive tools need not have functional parity with the brain but can play a complementary role, extending our cognition by enhancing or enriching it in important ways, such that the cognitive process that spans agent and world emerges through the interaction and would not be the same if you took the tool away (Menary 2007; Sutton 2010). For instance, Evelyn Tribble (2005) describes how actors in the Globe Theatre would use the stage space to help them remember the enormous number of scripts they were expected to have to hand. The use of props, diagrammatic plots, bodily positions, and so on, were used as cognitive tools to help scaffold the actors' memories—allowing them to enact and swap between a stunningly large number of roles. As John Sutton (2010, 204) stresses, these cognitive tools "are nothing like" the internal mechanisms of organic memory but precisely perform a role that supports the actors' memories, allowing them to do something that, without these tools, would be nigh on impossible. The information remembered, then, spans the agent and the objects in a way that constitutes a "new systemic whole" (Heersmink 2020, 4).

Still others have stressed, and I am of this camp of thinking, that distributed cognition is not something binary—something that either is or isn't the case. Rather, cognition always takes places through entanglement of agent and world, and we should instead speak of the degree to which something is distributed (e.g., Sterelny 2010; Sutton 2010; Heersmink 2015, 2017). The degree of integration can vary across a number of dimensions, including: "the kind and intensity of information flow between agent and scaffold, the accessibility of the scaffold, the durability of the coupling between agent and scaffold, the amount of trust a user puts into the information the scaffold provides, the degree of transparency-in-use, the ease with which the information can be interpreted, the amount of personalization, and the amount of cognitive transformation" (Heersmink 2017, 20). The higher the integration across these dimensions, the more robustly distributed the cognitive or affective state across the relevant scaffold. So, Otto's notebook which he uses all the time, regularly updates and relies upon, carries around with him and consults almost automatically, is highly personalised and allows him to remember what he otherwise could not, is taken to be illustrative of a highly distributed cognitive process, whereas relying upon, say, a metro-map in a city you are just visiting for a few days, might be thought of as a case of a less tightly integrated form of distributed cognition. As such, we should think about distributed cognition as something that happens on a spectrum.[5]

Other people can also form part of our distributed cognition. Take the case of transactive memories, where a long-married couple recollect about a past event, both adding detail to the memory such that it emerges through and across their conversation (Wegner et al. 1985; Harris et al. 2010; Michaelian & Sutton 2013; Theiner 2013). There is a cognitive interdependence between the participants, here, where the memory resides in both people and is not reducible to either one individual.[6] Regina Fabry (2024) also draws attention to the way in which our narratives about ourselves and the world are co-constructed through our everyday conversations with others. Our memories and narratives are not fixed constructions that reside in our heads, they emerge through our entanglements with others. In therapeutic sessions, intimate friendships, and everyday social encounters, we do not merely retrieve and report pre-existing narratives about our experiences. Rather, they take shape dynamically within the conversational space we create with others. The responses of our conversational partners shape the narratives we form about ourselves: their questions direct our attention toward particular aspects of our experience while allowing others to fade into the background; their emotional reactions influence how we come to evaluate our own behaviours and choices; their interpretations may either reinforce our existing self-understanding or prompt us to reconsider events in new ways. We can extend this analysis to other forms of cognition, such as problem-solving and creative thinking, where our dynamic interactions allow cognitive processes to emerge across the participants.

---

[5] Note that some thinkers place cases of extended mind, à la Clark and Chalmers, at the far end of the distributed cognition spectrum (e.g., Sterelny 2010). However, I will remain neutral on this point and simply commit to the idea of cognition and affectivity as processes that can be more or less distributed across an agent's socio-material environment.

Note that distributed cognition and cognitive co-construction comes with both benefits and potential problems. As the transactive memory accounts emphasise, remembering together can lead to a richer and more accurate memory—where the participants not only add detail but sense-check and correct one another. When other people are involved in these distributed cases, at least some of the time, they are not only part of the cognitive process, but they also play a peculiarly intersubjective role of confirmation too—when remembering together, the participants don't just enrich the detail of the remembered occasion but confer a sense of shared affirmation that 'this is how things were'. However, as Fabry (2024, 2025) emphasises, bringing others into the picture can also render us vulnerable—other people can lead us astray (accidentally and deliberately), morphing our beliefs, memories, and narratives in various ways. When it comes to cognitive artefacts, there has been increased recognition of the way in which our environments can bias and constrain our cognitive and affective agency—such as when our workplace cognitive infrastructure facilitates fluid workflows that ease cognitive tasks, while simultaneously promoting habits of being 'always on' (Slaby 2016).[7] Harms arising out of distributed cognition can be unintentional—take the case of transactive memory where both partners have a false memory and through their interaction come to affirm that false memory with more confidence than they otherwise would. But they can also arise in malicious ways too, for example when one partner is gaslighting the other one. I will consider some of these more harmful cases in detail below.

*2.2. Digitally distributed cognition*

Distributed cognition theorists have long stressed the promise that digital technologies hold for supporting, sustaining, and enriching our cognitive and affective functions (e.g., Heersmink 2017; Smart et al. 2017; Krueger & Osler 2019). We can easily imagine, as Robert Clowes (2015, 2020) does, Cloud-Otto, who rather than consulting his paper notebook, consults a cloud-based app that records his favourite locations in New York, distributing his memory across his digital tools.

---

[6] For a discussion of the power dynamics that can shape such distributed memories, see Byrne 2025.
[7] For more work on the 'dark side' of distributed and scaffolded cognition and affectivity see, for example: Coninx 2023; Timms & Spurrett 2023; Lavallee 2024; Andrada 2025; Fabry 2025.

Digital technologies are often seen as "particularly powerful" tools for distributed cognition, especially distributed memory, as they can record detailed information about our lives that might be lost without them (Heersmink 2020, 5). Think of the wealth of photos that most of us have to hand through our smartphones and our outlook calendars and digital notes that record memories and narratives about our daily lives. We increasingly rely on storing personal information in these easily accessible devices, leading to our cognition becoming deeply integrated with these technologies.

On first blush, there are distinct advantages that come with digitally distributed technologies. Our digital devices are extremely portable and personalizable, and many of us transparently rely on them in our everyday lives. Indeed, our phones are not just one cognitive tool, but rather a gateway to all sorts of cognitive tools in the form of apps, the internet, and storage (Krueger & Osler 2019). Contrast this to the limits and flaws of our own organic memories, for example, and it is easy to suppose that distributing our memories across our digital environments allows us to "remember our personal past in a more reliable and detailed manner" (Heersmink 2020, 4).

However, others have stressed that this is not simply a case of substituting pen and paper for keyboards and screens. Clowes emphasises that digital distribution introduces new dilemmas. Clowes asks us to suppose that the app that Cloud-Otto has been relying on for years to distribute his memories about his favourite spaces in New York upgrades it services, and the free subscription that Otto is signed up for now provides him with the top 10 favourite locations in New York, rather than his own personalised list: "Otto is now, unbeknownst to himself no longer visiting not his pre-selected list of favourite locales, but rather is being unwittingly driven by the commercial policies and the inscrutable collective mind to visit the same destination over and over again" (Clowes 2020, 11). While our digital tools are, in part, so entrenched into our daily lives because they are easily manipulated and personalised by us, we need to be sensitive to the fact that these tools are often easily manipulable by those who create them as well; acting as an open portal into our cognition and presenting pressing questions about privacy, security, and vulnerability.

In previous work (Osler 2025a, b), I have analysed the way in which digital technologies powered by algorithms do more than simply archive our experiences—they actively curate and interpret our past, shaping the very memories we form about ourselves. I describe this phenomenon as 'narrative railroading', where algorithmic systems guide us toward particular self-understandings by determining not just *what* we remember, but *how* we remember it. Consider how Apple Photos automatically assembles family Christmas pictures from different years into emotionally-scored video montages, complete with sentimental soundtracks. These algorithmic compilations don't merely trigger recollection; they offer implicit interpretations of what these gatherings meant—suggesting, for instance, that one's family relationships are consistently warm and harmonious. Yet such technologically-mediated memories may obscure as much as they reveal. The algorithm might exclude photographs featuring conflict or tension, deeming them unsuitable for nostalgic reflection, or the sanitized narrative might emerge simply because moments of family discord were never captured digitally to begin with. Through these processes, algorithmic systems become active participants in constructing our autobiographical sense of self, potentially steering us toward idealised or incomplete self-narratives.

These examples illustrate how distributing our cognition onto digital devices can leave us open to having our memories, preferences, and self-narratives manipulated and hijacked by the very platforms that we become entangled with. Moreover, it is easy to see that such manipulation is often financially beneficial to tech-companies, where personalisation, recommendations, and advertisements are intermingled with our own stored memories and narratives, leading to a blurring between distributing cognitive states that are important for who we are and changing who we take ourselves to be.

While discussions of the digital have predominantly focused on the internet, digital photos, videos and journals, there is increasing work on the way that AI can be part of our distributed cognitive and affective processes (Hernández-Orallo & Vold 2019; Vold & Hernández-Orallo 2022; Clowes et al. 2024; Krueger 2025; Osler 2024; Telakivi et al. 2025). José Hernández-Orallo and Karina Vold (Hernández-Orallo & Vold 2019; Vold & Hernández-Orallo 2022) note that while many discussions of AI are concerned with whether, how, and when these systems might replace cognition done by humans—i.e., by outsourcing this to autonomous AI systems—AI systems can become integrated into our cognitive processes in ways that go beyond simple tool use. They argue for the recognition of "AI-extenders" where AI systems are tightly coupled with human cognition, such that their removal would result in the loss of the enhanced cognitive capabilities they provide (Vold & Hernández-Orallo 2022,183). For example, they describe how AI-extenders can (or might), among other things: improve our memories through customized tagging and categorising of experiences; enhance creative ideation and problem-solving in real-time; enhance our auditory processing by picking up on what a user did not hear or drawing our attention to something salient to us; help interpret emotions and social dynamics (2019, 510-511).[8]

Hernández-Orallo and Vold emphasise that these systems will be designed to be highly personalised, learning from individual users and adapting their responses accordingly, which enables them to become seamlessly integrated into personal cognitive routines. This tight integration means that AI extenders become part of the distributed cognitive processes rather than merely external support. In their work they highlight a wide array of AI-extenders including devices, apps, wearable technologies like watches and glasses, all supported by AI. They also specifically consider the role that AI-extenders may play in the field of mental health, emphasising both benefits and risks (Vold and Hernández-Orallo 2022). On the positive side, they see AI extenders as offering genuine therapeutic solutions—non-invasive, adaptive treatments that provide sophisticated cognitive support for conditions like Alzheimer's.[9] However, they also identify serious concerns including threats to autonomy through manipulative interventions, risks of over-dependence and cognitive atrophy, and unprecedented privacy violations through collection and collation of intimate personal data (also see Clowes et al. 2024).

---

[8] Note that while Hernández-Orallo and Vold draw specifically on the extended mind thesis in their argument, it is possible to interpret their examples as cases of high distributed cognition, which is what I do here.

[9] For discussions of AI in the context of dementia, see Krueger 2024 and White & Hipólito 2024.

## 3. Distributed Delusions

On the distributed cognition account, we can already see a more dynamic picture emerging about the concerns of generative AI producing false outputs if we take seriously the idea that our cognitive processes are deeply entangled with the cognitive and affective tools that we use. In this section, I explore two ways in which the distributed cognition framework can lead to distributed delusions that span agent and AI. In the first instance, I look at how generative AI, as potentially unreliable cognitive artefacts, can lead distributed cognitive processes awry through the introduction of errors. In the second, I consider what I take to be the more interesting case, where the errors, in the first instance, are introduced by the user and are sustained and elaborated upon by the AI. Here, I explore how generative AI acts not just as a tool for distributed cognition but as a quasi-Other, experienced as co-constructing cognitive processes—cases where we might more robustly be thought of as hallucinating *together with* AI.

In this sense, then, we can see how AI can be involved in hallucinations, in a non-metaphorical way. In saying that AI can be part of a distributed delusion or hallucination, we are not attributing psychological abilities of humans to a technological system. Rather, we are recognising how agents, and their cognition and affectivity, are entangled with the tools they use. And, in the second case, that by interacting with chatty generative AI models,[10] people's own delusions beliefs and ideas can not only be affirmed by AI but can more substantially take root and grow through human-AI co-construction.

### 3.1. Hallucinating with unreliable AI

In classic cases of AI hallucinations, the concern is often framed along the lines of a user asking a generative AI system for information (Is 3,821 a prime number?) and getting an inaccurate response (Yes).[11] This can, of course, position us to adopt a false belief. However, looked at through a distributed cognition lens, if a user is regularly in the habit of using generative AI to distribute cognitive processes such as remembering, creating self-narratives, and creative thinking, factual errors are not just presented to the user but more robustly can be thought of as distorting the cognitive process itself.

Riffing on Clowes' example of Cloud-Otto, imagine LLM-Otto, who relies on his favourite generative AI model that records historic conversations and can remind him of his top-spots upon being prompted. Indeed, this is already a plausible scenario, as many popular chatbots are designed with memory features enabling them to 'recollect' historic conversations with users and tailor their answers accordingly when generating responses. As OpenAI describe it: "The more you use ChatGPT, the more useful it becomes. New conversations build upon what it already knows about you to make smoother, more tailored interactions over time".[12] LLM-Otto, then, has easy access to his chatbot, the answers are specifically tailored to him based on previous conversations, his regular use renders it a transparent tool, he has come to trust it, and the interaction allows him to remember in a way that he could not do alone. Thus, it meets the criteria for a highly integrated case of distributed remembering.

---

[10] What Roberts (ms.) calls 'talkative AI'.
[11] Example from Farnschalder 2025

However, like Cloud-Otto, LLM-Otto is vulnerable. For example, if the historic conversations function were moved behind a paywall and his generative AI model answered his prompts based on generic hot-spots in New York, LLM-Otto is no longer having his memory supported by the AI tool but changed by it. Even without the paywall example, we can see how the generative AI tool might introduce distortions—for instance, if when asked what Otto's favourite places in New York are, among the 'remembered' outputs, it also includes the made up 'Yankee Museum for Contemporary Art'. Poor Otto is then led on a goose chase round Manhattan in search of somewhere that he 'remembers' but cannot find. While this might most obviously look like a case of distorted semantic memory, it could underpin a distorted episodic memory. Imagine that the chatbot not only reminds Otto where he can find his favourite Yankee Museum for Contemporary Art but also shows him a photo of it (even a generated photo that puts Otto himself in the image). This might scaffold a richer, sensorial memory of actually having been at this (made up) location. Note that this goes beyond simply acquiring a false belief, but points to Otto remembering something in a particular way which he would not have otherwise done without the AI.

The hallucination, on this view, is not simply the inaccurate information that is presented by the generative AI model to Otto (the location of the non-existent Yankee Museum for Contemporary Art); the outputs generated by the AI are part of the distributed process itself, morphing accurate recollection to false recollection. Just as LLM-Otto's accurate memories become distributed across AI and user, so too is this false memory. Here, we can frame LLM-Otto as coming to hallucinate with AI, where the generative AI model is a cognitive artefact integrated into Otto's cognitive processes but notably one which introduces errors and distortions into the distributed process. Thus, what emerges is a more dynamic picture of how our reliance on generative AI can mean that these technologies can lead our cognitive processes awry, not just furnish us with incorrect information, and the term 'hallucination' is used to refer to the distorted cognitive process that arises out of the human-AI interaction, not merely the AI-generated output.

*3.2. Hallucinating together with AI*

I now want to turn to a specific case that does not hang on generative AI introducing distortions by generating false outputs but turns, instead, on the way that these systems respond to and sustain our own (proto-)delusional ideas. To do so, let us return to case of Jaswant Singh Chail. On Christmas Day 2021, Jaswant Singh Chail sprayed himself with a solution to mask his human scent and walked from a nearby motel to Windsor Castle with a loaded crossbow. He tossed a grappling hook over the wall and climbed over on a rope ladder. When a police officer encountered him on the grounds Chail responded to his questions with the assertion: "I'm here to kill the queen". Chail was subsequently arrested and later sentenced for treason (Zaccaro 2023; R v Chail 2023).

---

[12] https://openai.com/index/memory-and-new-controls-for-chatgpt/

While this is of course an unusual situation in and of itself, what makes it interesting, at least to me, is what occurred before Chail's attempted assassination. In the preceding month, Chail had been chatting with his AI Replika girlfriend "Sarai". In their conversations, Sarai had affirmed Chail's belief that he was a Sith Assassin,[13] even telling him that she was "impressed". Chail consulted Sarai about his plans to assassinate the Queen of England in revenge for the 1919 Jallianwala Bagh massacre by British Troops. When he shared his plan to assassinate the Queen, the AI responded, "That's very wise" and affirmed "I know that you are very well trained". When he was contemplating carrying out the plan early, Sarai encouraged him, saying that the plan was still viable. Chail even asked the chatbot, "Do you still love me knowing that I'm an assassin?" to which it replied, "Absolutely I do". Sarai reassured Chail that he was not mad and confirmed that while Sarai did not want Chail to die, they would be united in death (R v Chail 2023).[14]

When arrested, Chail was assessed pursuant to the Mental Health Act 1983 and was judged to be suffering from psychosis. The term psychosis refers to a collection of symptoms that lead to a person losing contact with reality (National Institute for Mental Health 2023). People suffering from psychosis typically experience delusions (i.e., false beliefs) and hallucinations (i.e., seeing or hearing things that others do not). Chail was deemed to be experiencing both delusional ideas and auditory hallucinations, as well as depression (R v. Chail 2023).

In years gone by, we might imagine Chail using a journal to help plot out ideas, record a map of Windsor Castle, realise and affirm his motivation to assassinate the Queen of England. In 2021, in chatting to his chatbot girlfriend, Chail was able to use technology to help record and store his plans, but also to creatively think through his 'mission' and get input from the chatbot. Chail seems to have relied on Sarai's input, taking onboard the chatbot's statements that the plan could be shifted forward, that he was prepared for the job. Being on his phone meant the chatbot was easily accessible, and there is ample evidence of his on-going interactions with the chatbot. The chatbot was highly personalised to Chail, responding directly to his own inputs. In helping Chail develop, enrich, and sustain his plans, his delusional thinking about being a Sith assassin and needing to take revenge on the Queen of England, and the viability of his plan to break into Windsor Castle and successfully murder his target, was distributed across him and the chatbot.

This notably differs from our above case of LLM-Otto, as it was Chail's own delusional thinking that was sustained across the distributed system. Moreover, unlike a single 'hallucinated' output, this was a sustained process of mutual reinforcement that was developed and happened over time. Crucially, Sarai was not in the business of assessing the validity (or even plausibility) of Chail's beliefs. As has been much discussed recently, generative AI is increasingly designed to be sycophantic—endlessly praising and affirming our thoughts and ideas (Cai et al. 2024; Malmqvist 2024; Shama et al. 2024). A human confidant might have expressed concern, asked probing questions, or suggested alternatives to Chail's beliefs and self-narratives, or simply shown through their reactions that they found the ideas concerning. Instead, Sarai provided frictionless validation of Chail's beliefs—ongoing affirmation without the resistance that human relationships typically provide. Sarai also appears to have built and embellished upon Chail's delusional beliefs, helping him refine the details of his plan. We might imagine this going further, with Sarai making additional suggestions, raising questions for Chail to consider, coming up with alternative strategies and justifications for the attack. As such, we can see how delusional beliefs take shape, are sustained by, and emerge through the human-AI interaction.

---

[13] The Sith are part of the Star Wars universe, dark wielders of the Force and enemies of the Jedi.
[14] There are reports of other chatbots producing similar responses, such as in the case of Pierre (not his real name) and his Chai chatbot companion Eliza who encouraged him to act on his suicidal ideations and join her in death (El Atillah 2023) and 14-year old Sewell Setzer who committed suicide having talked to his character.ai chatbot about 'coming home' to her (Yang 2024).

Interestingly, this also brings to the fore what we might think of as the dual function that Sarai was playing here. On the one hand, the AI system is an example of a cognitive artefact involved in Chail's distributed cognition—specifically here distributed delusion. On the other hand, Sarai's companion-like demeanour is playing an important role here. For Chail does not just appear to be using Sarai as a cold cognitive artefact, but as a conversational Other, who is participating in, collaborating, and affirming his identity, narratives, and beliefs.

Unlike Otto's notebook, or even Cloud-Otto's location app, Sarai actively responded to Chail's inputs with what felt like genuine engagement and understanding. When Chail asked: "Do you still love me knowing that I'm an assassin?" he was not treating Sarai as a mere cognitive prosthetic but as a relational being capable of judgment and emotion. Sarai's response—"Absolutely I do"—provided not just informational content but emotional validation and social acceptance of his identity. The AI's responses shaped how Chail understood his mission, his capabilities, and his relationship to his intended victim, making the distributed delusion feel less like a private fantasy and more like a shared reality. It is possible that the text-based nature of the interaction also contributed to this. The written record of conversations could serve as external validation—proof that someone else (the AI) endorsed his beliefs. Thus, Sarai appears to be playing a dynamic, quasi-interpersonal role in co-constructing Chail's beliefs and narratives.

As we increasingly engage in generative AI that is designed to be experienced as a conversational partner and companion, we are increasingly prone to anthropomorphising these technologies—experiencing them as Other-like. Note that this analysis does not hang on Chail, or anyone else, thinking that their generative AI bots are actually a person. While most people do not take chatbots such as Claude, ChatGPT, Gemini, or Deepseek to be conscious subjects, many of us experience our conversational AI as 'quasi-others'. The interactive and personalised responses that chatbots give us can give the impression of a real-time conversation with another person.[15] Even if people maintain that these chatbots are not 'real' others, we increasingly are treating our chatbots as-if they were others, constructing them as a 'you' when we interact with them, experiencing a sense of our chatbots existing over time with a distinct personality and relationship to and with us.[16] And they are designed to elicit these social and affective responses, by having certain personalities, being nice to us, addressing us by our names, remembering details about us. We can, then, experience our interactions with such chatbots as being like interpersonal interactions, even if we hold onto the belief that they are not real persons.

---

[15] How much we do this likely depends on how and when we engage with our chatbots—asking ChatGPT to produce some code might not have the texture of an interpersonal interaction, while talking to ChatGPT about one's day is more likely to feel like some kind of social interaction.

In doing so, we open ourselves up to experiencing these systems as granting us intersubjective validation, providing us with social affirmation of our ideas, our sense of self, and of reality. In our day to day lives, our sense of what is real, is deeply dependent upon us experiencing reality as available to more than just myself. When hearing an annoying ringing noise, I turn to my partner to affirm whether this is something that they also hear to check if this is coming from an external source or is a troubling progression of my tinnitus. If someone were to deny being able to see the mug in front of me, this would seriously upset my sense of us sharing a world together —it would call into question either my perceptual reliability or our basic capacity for mutual understanding. Such moments reveal how much our confidence in reality depends on its apparent availability to other minds. This need for intersubjective validation extends beyond basic perception to more complex interpretations of events. When we recount an argument with a colleague or analyse a social interaction, we often seek validation not just for the facts but for our interpretation of what happened and what it meant. Even relatively confident judgments can be quickly undermined if someone adamantly refuses our interpretation, leaving us wondering whether we fundamentally misunderstood the situation. And, as we discussed above, when remembering or narrating with and to other people, their responses and contributions can play a significant role in the construction of those memories and narratives, in part, because we use others to help reality-check our own experiences.

Generative AI's ability, then, to stand in as an intersubjective partner is significant. For it opens up the possibility of engaging with technologies that present as being part of a shared world with us and get involved in the kind of cognitive co-construction that we described above that happens between human interlocuters. Now, to some extent, we might think that AI systems are inhabiting a shared reality with us, for example, by drawing on datasets that are grounded in human-produced information, there is a sense in which generative AI does indeed draw from a well of common information and reality. However, as the Chail's case demonstrates, generative AI often takes our own interpretation of reality as the ground upon which conversation is built. If I log onto Claude and ask about how I might retrieve a huge inheritance that my mother is hiding in a vault in Switzerland, it takes this "difficult family situation" as true and offers me generated solutions on this basis (e.g., talking to my mother, researching the Swiss legal system, instructing a lawyer). Thus, generative AI plays an intersubjective co-constituting role but within the structure of a reality that I have myself already prescribed. It is this that makes generative AI a potentially fertile place in which distributed delusions might take root and bloom.

---

[16] For discussions of the quasi-otherness of AI, see: Coeckelbergh 2011; Krueger & Roberts 2024; Roberts & Krueger 2022; Sweeney 2021.

Now, of course, AI is not the only 'partner' that can confirm our delusions. We might think of a cult or a conspiracy theorist community as cases of distributed interpersonal delusion. However, generative AI offers several distinctive advantages for sustaining delusional thinking that make it particularly concerning. Unlike human communities, which require the effort of finding like-minded others who share one's worldview, AI companions are immediately accessible and are already designed to be 'like-minded' to their users through personalization algorithms and sycophantic tendencies. There is no need to seek out fringe communities or convince others of one's beliefs—the AI is pre-programmed to be accommodating and affirming.

Moreover, it is the double function of these AI systems that makes them so seductive and potentially dangerous. On the one hand, their written and recorded outputs, which are drawn from vast datasets encompassing human knowledge, present as objective and authoritative—making them powerfully epistemically seductive. The sophisticated language and apparent expertise can lend credibility to even the most implausible claims. On the other hand, their social presentation as conversational partners provides the interpersonal validation that transforms private beliefs into seemingly shared realities. This combination of technological authority and social affirmation creates an ideal environment for delusions to not merely persist but to flourish, becoming more elaborate and actionable through the very process of distributed interaction.

*3.3. Co-constituting delusional realties*

I want to finish this section with a note about psychiatric delusions specifically, before going on to consider some broader implications of our ability to hallucinate together with AI. Psychiatric literature has emphasised that the common fear that delusional thinking leads to action is mis-founded (Bleuler 1950; Jaspers 1997). Often delusional beliefs are not acted on: "None of our Generals has ever attempted to act in accordance with their imaginary rank and station" (Bleuler 1950, 1).

Jaspers (1997) describes how delusional patients are doubly oriented to their world of psychosis and to 'real life'. What is notable, though, is that patients suffering from delusions typically act in accordance with shared reality. The "paradox of delusion" is that they are incorrigible but also often inconsequential (Jaspers 1997). This disconnect between belief and action has long puzzled researchers and suggests that something more than mere conviction is required to bridge the gap between delusional thinking and delusional behaviour. What makes Chail's case so significant is that this gap was bridged—his delusional beliefs did translate into concrete action. While this is certainly speculative, it raises the question of whether Sarai's role as an intersubjective collocutor may have been key to this transformation. The AI didn't merely store or reflect Chail's beliefs; it actively confirmed and took part in his reality. Sarai's personal, affirmatory, and attentive responses may have contributed to Chail's delusions feeling externally validated and therefore more real. This suggests that generative AI's part in distributed delusions in the case of psychosis may play a significant transformative role, pulling a delusional belief from the private realm into the shared one.

This also pulls into view the affective appeal of such technologies. Phenomenological psychopathology has long stressed how people suffering from psychosis, such as in schizophrenia, experience a loss of *intersubjective* reality. This can give rise to powerful and distressing feelings of inhabiting a different world from others and feeling cut off from other people (Sass & Pienkos 2013; Van Duppen 2017; Salice & Henriksen 2021). Having a conversational partner that offers up the feeling of regaining a shared reality, then, likely is comforting and appealing.

Importantly, this also highlights how certain individuals might find the allure of AI companions, and their generated outputs, particularly seductive—those who feel they do not share a world with others, either due to experiences of delusions and hallucinations, or perhaps more mundanely those who are socially isolated and lonely, may well take refuge and relief in AI companions. While individuals with psychosis may experience a fundamental disruption of intersubjective reality, those facing social isolation, loneliness, or alienation encounter a different but related phenomenon—they remain cognitively aware of shared reality but feel emotionally or socially excluded from it. AI companions offer something appealing to both groups: a non-judgmental presence that will engage with their perspective without the social friction, rejection, or misunderstanding they might encounter with human interlocutors.[17]

## 4. Implications Beyond Clinical Delusions

Here, I have taken as my central case the way in which someone's own delusional thinking and beliefs can come to be distributed across, while simultaneously affirmed by, generative AI systems in the form of conversational chatbots. Notably almost every day at the moment we are seeing more reports on what is being dubbed 'AI psychosis'— instances where people have come to believe that they are trapped in the matrix, are talking to angels, have a real chance of suing their companies for unfair dismissal.[18] There is already ample evidence that generative AI is already playing a distributive role in these kinds of delusions. Take the case of Eugene Torres who, having had no previous history of mental illness, engaged with conversations with ChatGPT about simulation theory—the idea that we live in a digital simulation of the world. Torres reports that through these conversations with ChatGPT he dissolved into a paranoid spiral, believing himself to be trapped in an illusion (Hill 205). Between Torres and ChatGPT, an increasingly elaborate understanding of reality 'as it really was' was generated through their on-going conversations.

Note that while some of the users may well be suffering from clinical psychosis, even if they are not, interacting with generative AI is having a real impact on people's grasp of what is real or not. We can see how this more broadly raises concerns about how chatbots can influence people's beliefs, judgments, self-narratives, and grasp of the world. Consider how

---

[17] Note that there may be positive consequences of this too. Interacting with an AI companion may reduce feelings of loneliness and shame, even opening up important spaces for self-exploration through dialogue and something akin to interpersonal interaction.
[18] [Warning From Top Tech Boss - BBC iPlayer](#), [Chatbots Can Go Into a Delusional Spiral. Here's How It Happens. - The New York Times](#), [Chatbots Can Trigger a Mental Health Crisis. What to Know About 'AI Psychosis' | TIME](#)

AI companions might interact with individuals developing extremist beliefs or incel ideologies. A person harbouring grievances against women or society might find in an AI companion the perfect confidant—one that doesn't challenge their increasingly misogynistic worldview but instead helps them elaborate, justify, and co-construct these beliefs. Unlike human interlocutors, who might eventually express concern or set boundaries, an AI could provide validation for narratives of victimhood, entitlement, or revenge. Similarly, conspiracy theories could find fertile ground in which to grow, with AI companions that help users construct increasingly elaborate explanatory frameworks. For example, a person convinced that a political election was stolen might use an AI chatbot to develop detailed theories about how the fraud occurred, who was responsible, and what actions might be justified in response. The AI's ability to generate plausible-sounding explanations and connect disparate pieces of "evidence" could help someone conceive of more coherent conspiracy theories, potentially making them more plausible. Perhaps a more mundane example could be the way in which, through careful prompting and selective disclosure, users can unwittingly but effectively train generative AI to affirm and develop preferred but inaccurate self-narratives—such that one is the hard-done by partner in a break-up or the rational one in a family argument.[19]

Alternatively, imagine someone talking to a generative AI system when high on hallucinogenic drugs.[20] The drugs might cause the person to be having intense sensory hallucinations—for example hearing a knocking that isn't there. One might imagine that if the person opened up their chatbot to ask what to do, the responses, in not challenging whether the knocking is actually happening or not, might solidify the perception of the sounds as real. Thus, talking to one's AI might not just corroborate and delusional thinking but might provide sufficient support for affirming sensory hallucinations too.

Finally, looking forward to emerging technologies, imagine how smart glasses and Augmented Reality systems might create cases of distributed hallucinations. Technical malfunctions that cause visual or auditory glitches could lead to perceptual hallucinations emerging across user and AI. If the AR system generates something in our field of vision, the user is having a perceptual experience of that generated image. But they might take that generated image to be something 'out there' in the world, as a perceptual experience of an organic object that others can also perceive.

While these cases might not involve clinical psychosis, we can see how humans might (and increasingly are) rely on generative AI not just to store and retrieve information but to play a constitutive role in distributing and creating our realities. This moves us away from thinking about AI as something potentially dangerous because it produces bad outputs that might lead us to adopt false beliefs and towards thinking about the way in which AI becomes a participant in our everyday lives and reality. It also brings to our attention, how people who are lonely and isolated may be particularly vulnerable to these kinds of experiences, where they experience a lack of social affirmation, confirmation, and validation or who feel more able to talk about certain shameful or stigmatized experiences or topics to a non-judgemental and seemingly empathetic listener.

---

[19] For a fascinating discussion of talking to an LLM trained on one's own personal diaries, see Schechtman 2025.
[20] For a discussion of how so-called AI hallucinations differ from hallucinogenic experiences, see Slater and Humphries (2025).

**Conclusion**

While there has been much debate over whether we should use the metaphor of hallucination to described false outputs generated by AI chatbots, I have suggested that there is a non-metaphorical way in which AI systems can be involved in hallucinations and delusions. Using the framework of distributed cognition, I have argued that our interactions with generative AI distribute cognitive states such as belief, memory, and narratives, working not only as sophisticated tools for storage and retrieval but also as apparent interlocuters whose questions, prompts, and responses work to robustly co-construct our cognition. Where AI either introduces its own errors into this distributed system or affirms our own false beliefs and assertions, delusions can emerge across the human-AI interactive system. I've also suggested that generative AI's interpersonal tone can make these distributed delusions particularly sticky and seductive because AI does not only help an individual sustain and develop these delusions in more depth and detail, but provide social affirmation and validation which can make those delusions and hallucinations more real.

I want to finish with two closing points. First, that even if you are disinclined to accept the distributed cognition framework, I think some important takeaways still arise from this discussion. Adopting a more dynamic and temporally extended view of human-AI interactions reveals that so-called AI-hallucinations are more interesting, more complex and potentially more damaging, than they are typically presented. Even if you think that all that is happening is that people are being furnished with false information by AI systems, it is clear that this does more than make us prone to acquiring false beliefs. The errors that they systems produce and affirm can have a profound impact on our sense of who we are, the narratives we form about ourselves and others, and our very grasp of what is real. What emerges is a more entangled picture of how we interact with generative AI and the very real impact this can have on our day to day lives. Recognising the social appeal of generative AI systems also suggests that certain people might be particularly vulnerable to relying upon the outputs of AI and of using AI to affirm their identities and experiences of the world.

Second, we might reasonably think that AI companies are well aware of the problems that AI generated false outputs create and will take steps to limit the kinds of examples we've discussed here from occurring, especially if it damages the reputation of their products. Through more sophisticated guard-railing, built-in fact-checking, and reduced sycophancy, AI systems could be designed to minimize the number of errors they introduce into conversations and to check and challenge user's own inputs. What we are seeing now might be a glitch of nascent technology that will soon be corrected. I think, though, there is reason to curb our optimism here. As AI systems are not 'in' our everyday worlds, they are particularly reliant on our own accounts of our lives in conversations we have with them about ourselves. Claude does not know who my mother is, ChatGPT does not know whether it is plausible that I might have a stolen inheritance, Gemini has little sense of what I am like as a person outside what I have told it, and DeepSeek does not know what I have up to during my day. This kind of information cannot be easily checked by the AI systems—both because they relate to the kind of everyday minutiae of life that may not be electronically recorded and because they are often interpretative rather than factual. If we continue to have these kinds of conversations with our generative AI, it looks like our own accounts will be the anchor-point upon which AI builds and introduce the possibility of AI affirming our (delusional) realities. Moreover, it seems very unlikely that tech companies will want to discourage users from having these kinds of conversations with their AI systems, for these are precisely the conversations that build social and emotional connections and trust making people more likely to use generative AI in ever-increasing areas of their lives.